\documentstyle[12pt,aasms4]{article}

\lefthead{Pizagno and Rix}
\righthead{The Extinction Distribution in the Galaxy UGC~5041}

\def\plotone#1{}

\def\gtorder{\mathrel{\raise.3ex\hbox{$>$}\mkern-14mu\lower0.6ex\hbox{$\sim$}}}
\def\ltorder{\mathrel{\raise.3ex\hbox{$<$}\mkern-14mu\lower0.6ex\hbox{$\sim$}}}

\newcommand\etal{{\it et al.~}}
\newcommand\eg{e.g.,~}

\newcommand\asec{^{\prime\prime}}

\begin{document}

\title{The Extinction Distribution in the Galaxy UGC~5041}

\author{James Pizagno,\altaffilmark{1}
        Hans-Walter Rix\altaffilmark{1,2}
        }

\altaffiltext{1}{Steward Observatory, University of Arizona, Tucson, 
                 AZ 85721}
\altaffiltext{2}{Alfred P.~Sloan Fellow}

\def\arc{$'$$'$}

\begin{abstract}

We probe the dust extinction through the foreground disk of 
the overlapping galaxy pair UGC 5041 by analyzing 
B,I, and H band images. The inclined foreground disk of this 
infrared-selected pair is almost opaque in B at a projected
distance of $\sim 8$~kpc. From the images, we estimate directly the
area-weighted 
distribution of differential near-IR extinction: it is nearly
Gaussian with $\langle \tau_I-\tau_H\rangle =0.6$ and
$\sigma=0.27$.
For a homogenous dust distribution and a Milky Way
extinction curve, this corresponds to a face-on
distribution p$(\tau)$ with a mean of $<\tau_V>=0.34$ 
and $\sigma_V=0.15$.  For a clumpy 
dust model the optical depth estimate increases to
$<\tau_V>=0.41$ and $\sigma_V=0.19$.
Even though the galaxy pair is subject to different selection biases
and our analysis is subject to different
systematics, the result is consistent with existing case
studies, indicating that $<\tau_V>\sim 0.3$ is generic for late-type
spirals near their half-light radii.

We outline how to estimate from p$(\tau)$ by how much background
quasars are underreresented, where projected within $\sim 10$kpc 
of nearby spirals,
such as damped Ly-$\alpha$ absorbers or gravitational lenses; from our 
data we derive a factor of two deficit for flux-limited, optical
surveys.
\end{abstract}

\keywords{galaxies: spiral --- galaxies: individual (UGC 5041) --- galaxies: ISM --- ISM: dust,extinction}

\section{Introduction}

The disks of spiral galaxies contain enough dust to 
affect significantly the UV and optical radiation from the stars
within the galaxy and from any sources behind them (see
\eg Davies and Burstein, 1992, for an overview).
The effect of dust extinction on the emerging spectral energy distribution
at $\lambda\le 2\mu$m can be subtle, even if large quantities
of dust are present (\eg Phillipps \etal 1992, 
Witt \etal 1992, Rix 1995), and depends 
strongly on the relative layering of the 
stars and the dust, and the small-scale clumping of the dust.
The impact of dust extinction 
on both the colors of the galaxy itself and on the statistical 
extinction and reddening of background sources depends
also on the small scale clumping of the dust (\eg Witt and Gordon, 
1996; Berlind \etal 1997,
hereafter B97)
and on its concentration towards the spiral arms (White and Keel, 1992,
hereafter WK92; Keel and White 1995).

Even though galaxy disks (and their associated dust)
may cover only a few percent of the sky to redshifts
$z\gtorder 1$, they affect particularly interesting lines-of-sight.
For example, the true rate of damped Ly-$\alpha$ systems
depends sensitively on the degree to which the detectability
of the QSO (usually in the optical) is affected by potential
dust extinction in the absorber (Bartelmann and Loeb, 1996).
Similarly, the incidence of multiple gravitational images and 
of lensing magnification caused by disk galaxies may be altered by
 the presence of dust that may prevent the cataloging of the source
(Rodrigues-Williams et.al. 1994; Boyle et.al. 1988).

Measuring the optical depths of spiral disks is most
straightforward with background light sources,
because the influence of dust scattering is minimized, leaving only
extinction. However, background point sources are insufficient
to study the spatial structure of dust extinction.
Keel (1983) and White and Keel (1992)
pioneered a technique to map the extinction
distribution in a galaxy, by analyzing the light distribution 
of a background galaxy shining through it.
Using three pairs of galaxies, they found that in these systems the 
face-on B-band extinction near the spiral arms was $\tau_B\sim 0.3$,
and $\tau_B\sim 0.075~-~0.2$ in the interarm-regions.

As the results of such an extinction analysis depend on the objects chosen,
the radial range probed, the available spatial resolution and the 
wavelengths of observation, it is important to increase the number of 
case studies and employed techniques to arrive at a coherent and comprehensive
picture of dust extinction in spiral galaxies.
Here we present an analysis of the projected galaxy pair
UGC~5041, following the approach of WK92.
The forground galaxy is a highly inclined Sc at $cz=8083$~km/s,
with a background galaxy at $cz\approx 12,000$km/s, whose
center is projected at a mean distance of $14.4\asec$ (or $8h^{-1}_{70}~kpc$)
from UGC~5041's center.

We decided to present another case study, to complement the existing
small set of studies already in the literature (WK92 and B97),
because the particular galaxy pair, the 
observations and analysis differ in several relevant aspects:
\begin{itemize}
\item The background galaxy behind UGC~5041 was found only in H$(1.6\mu$m) images
(Gary Bernstein, {\it priv. conv.}); it is completely
obscured in the B-band. Hence, the galaxy pair is subject to different
selection biases than optically selected pairs and has the potential
to yield an optical depth estimate of $\langle \tau_B\rangle\gtorder 1$.

\item Given the relatively small projected separation ($R_{proj}\approx 1.4R_{exp}$)
and qualitatively high optical depths (see also NGC~3314, Keel 1983),
our analysis is based solely on 
near-IR observations ($I$ and $H$).

\item From the observations we calculate the {\it area-weighted optical depth
distibution}, p$(\tau)$, near the half-light radius of the foreground disk.
This is the most immediately relevant quantity for assessing
the extinction statistics of background sources.

\item We apply the derived p$(\tau)$ to explore the
under-representation of QSOs lensed by spiral galaxies 
and of QSOs with low redshift damped Ly-$\alpha$ absorption. 

\end{itemize}

The paper is organized as follows. In \S 2 we describe the
data reduction and analysis.
In \S 3 we describe the estimates of the optical depth
distribution and its errors; in \S4 we discuss 
the implications of this estimate and in \S 5 we present the
conclusions.

\section{Observations and Reductions}

\subsection{Observations and Initial Reduction}

The H-band (1.6$\mu$m) images were taken on September 19, 1996 at the
2.3m Bok telescope on Kitt Peak using the 256x256 IRcam with a scale
of $0.2''$/pixel.  Bias and dark frames were obtained at the beginning of
the night.  The data were taken by integrating on the object for 60
seconds, shifting to the sky ($\sim 5'$ off-set) for 60 seconds,
then returning to the galaxy and repeating the procedure. Each telescope 
dither was to a slightly different part of the sky near the object.  
The total integration time on the galaxy pair was 3600~sec    (60 object-sky 
image pairs). In the subsequent analysis, eleven of the
final images were discarded, because crucial portions of the
image were corrupted by bad pixels.
        
Most of the data reduction was carried out with IRAF\footnote{
IRAF is distributed by the National Optical Astronomy Observatories,
which are operated by AURA, Inc., under cooperative agreement 
with the National Science Foundation.}.  The bias
and dark frames were subtracted from all galaxy and sky frames.
Subsequently, the two sky flats bracketing an object image were averaged,
normalized and used to flatfield the interleafing galaxy image.
The resulting images were then aligned using the foreground
galaxy center and co-added.

Reduced and flux-calibrated I-band(7800\AA) and the B-band(4400\AA) CCD
images of UGC5041 were kindly provided by Gary Bernstein. They were
originally taken on February 27 and 29,
1992, respectively, at the 2.3m Bok telescope on Kitt Peak.
Since the optical images had different pixel scales (0.3$''$/pixel),
the H-band image  need to be de-magnified and aligned.
All photon-noise and flux-calibration errors were propagated throughout these
steps, but in the end found to be  negligible compared to the systematic
modeling uncertainties discussed below.

Figure 1 shows the resulting images in B, I and H.  It is 
apparent that in the
B-band the extinction is high, for the most part obscuring the background
galaxy. Only one bright region is shining through (in the upper left part
of the foreground galaxy), explaining why this galaxy pair was never found
in optical surveys\footnote{It was found serendipidously by G. Bernstein in
a low resolution H-band image.}.  Qualitatively,
the I-band image hints at $\tau_I\sim 1$,
as the dust extinction is obvious, but not overwhelming.
The rightmost panel of Figure 1 shows that dust extinction effects in the 
H-band are subtle.  However, we do not neglect the H-band extinction.

\subsection{Estimates of $\tau_I - \tau_H$}

With the three images aligned, scaled, and flux-calibrated we can
proceed to estimate the optical depths in the foreground galaxy.  The
recorded images are composed of light from the foreground galaxy and of
attenuated light from the background galaxy.  The equation for
radiative transfer in this case is simply (see also WK92)
\begin{equation}
 I^{\rm obs}_{\lambda}(x) = I^{\rm fore}_{\lambda}(x) +
    I^{\rm back}_{\lambda}(x)~e^{-\tau_{\lambda}(x)} 
\end{equation} 
where $\lambda$ denotes the wavelength of the observations, and x the
image position.  Since the extincted source is well behind the
extincting dust we neglect scattering into the
line of site.  From Eq.(1) one can see that as a first step
we must subtract the foreground galaxy light,
$I^{\rm fore}_{\lambda}(x)$, at each wavelength in the region
where the two galaxies overlap.  Since the edge-on
background galaxy is an almost linear feature, 
an estimate of the foreground galaxy light was made by 
interpolating the adjacent foreground light across it.
This fit was then subtracted from the overlapping
region,  and the associated errors were an
important  source of uncertainty
in the final optical depth determination. Indeed, the HII
regions and the spiral arms in the foreground galaxy rendered
the B-band image too irregular to get an accurate estimate for the fit.
Therefore, the B-band data was not used in the subsequent analysis.
The H-band image on the other hand, had little error in subtracting
out the foreground light, because the foreground galaxy contributed
little to the light in the overlapping region to begin with.  
To get an independent error estimate for
$I^{\rm obs}_{\lambda} - I^{\rm fore}_{\lambda}$ we
analyzed the opposite side of the foreground galaxy, where there are
no over-lapping galaxies, and carried out the same interpolation.
The residuals provide us with an accuracy estimate
of $\sim 6\%$ in I for the foreground subtraction.

Next the corrected background
light, $I^{\rm obs}_{\lambda}(x) - I^{\rm fore}_{\lambda}(x)$, should 
be divided by an estimate
of its {\it un}extincted value, $I^{\rm back}_{\lambda}(x)$,
to solve for $\tau_\lambda (x)$.
Although direct estimates of $I^{\rm back}_{\lambda}(x)$ are unknown, 
because we cannot neglect the foreground extinction even in the H-band.  
Fortunately, the intrinsic I-H color of highly inclined
disk galaxies is quite uniform. 
For example, Bernstein et al (1994) find I-H=1.88$\pm$0.08 magnitudes for their
sample of 21 spirals, with 2 outlier galaxies that are bluer by 0.4 
magnitudes than the mean.  Hence, we have a very good guess of the background galaxy's color
in absence of the foreground galaxy.  By dividing the observed I-H flux 
ratio by the un-extincted, or intrinsic, value,
we can estimate $\tau_I - \tau_H$ directly via
\begin{equation} 
\tau_I - \tau_H = -\ln\left({{I^{\rm back}_{I}} \over {I^{\rm back}_{H}}}
~/~ {{I^{\rm intr}_{I}} \over {I^{\rm intr}_{H}}}\right). 
\end{equation}
To obtain a well defined map of $\tau_I - \tau_H$ we created a mask
that isolated the image regions of significant background galaxy 
flux in H.
The pixel values in the masked ratio-image can then be converted into
a histogram of $\tau_I - \tau_H$, which is shown in Figure 2.  Our data span a 
range of $\sim3.5h^{-1}_{70}$~kpc in the plane of the foreground galaxy, but no 
systematic radial variation of the extinction was found.

Uncertainties in the optical depths were calculated by standard propagation of
errors (Bevington 1992), accounting 
for the foreground subtraction and the intrinsic color ratio variance.
Intrinsic color ratio errors contributed 0.1 to $\tau_I-\tau_H$,
 and the error contribution from the subtraction depended on 
 $I^{\rm obs}_{\lambda} - I^{\rm fore}_{\lambda}$.
The final median error in $\tau_I - \tau_H$ was found to be $\pm$0.13.

\section{Analysis}
\subsection{Estimating $p\bigl(\tau_V\bigr )$}

Our data provide a direct estimate of
the differential extinction probability distribution
$p\bigl(\tau_I - \tau_H\bigr )$.  For broader applications it would 
be useful to correct this value to face-on and to convert it
to estimates of $\tau_V$.  The radial range( $6.4$ to $10.3$~kpc) covered
by our estimate of p$(\tau_V)$ is quite limited, but of particular interest.  
First, because damped Ly-$\alpha$ absorbers have sizes $\sim10$~kpc (Briggs et. al. 1989).
Second, because the Einstein ring radius for distant galaxies is $\sim4$~kpc(e.g. Schneider et. al. 1992).
As a first step and benchmark, we estimate $\tau_V$ by assuming an
extinction curve, $\tau = \tau(\lambda)$, similar to the Milky
Way's, (e.g. Mathis, 1990). Under this assumption the face-on
value of  $\tau_V$ is given by 
\begin{equation}
\tau_v^{face-on} = \alpha \Bigl(\tau_I-\tau_H\Bigr )~\cos{i},
\end{equation}
where $\alpha\approx 2.15$ (Mathis 1990).
   The inclination angle, i, of the foreground galaxy was
found by fitting isophotes. Assuming an edge-on axis ratio of
0.19 for UGC~5041's stellar disk 
we find an inclination angle of $75^{\circ}\pm 2^{\circ}$;
the inclination error turned out to be negligible in the
subsequent analysis.

A pixel-to-pixel histogram of the face-on optical extinction, $\tau_V$, derived under
these assumptions is shown
in Figure 3.  The distribution of the face-on $\tau_V$ is approximately
Gaussian with $\langle \tau_V\rangle \approx 0.34$,
and a variance of $0.15$.

\subsection{Patchy Extinction Models}

In converting $\tau_I - \tau_H$ to $\tau_V$ we assumed so far that
the extinction curve in the foreground galaxy was similar to the Milky
Way's.  However, the IR extinction was derived over an extended region
of the foreground galaxy, as each $0.3\asec$ pixel subtends
$164~pc$ on a side at a distance of $114 h^{-1}_{70}$Mpc.
Casual inspection of nearby galaxy images
and quantitative studies (see \eg Witt and Gordon, 1996; 
WK92, B97) make it clear
that the dust is quite clumpy on much smaller scales.
In this section we investigate how much dust inhomogeneities
affect our estimates of p($\tau_V$) by using a simple model
to derive an effective
coefficient, $\alpha_{\rm eff}$, for clumpy extinction.

Following B97 we consider an area A in the
foreground galaxy, covered by one pixel, within which a fraction
$f$ is covered by material with an $X$ times greater extinction curve
than the rest of the area.  The incoming intensity $I_0$, emerges as
$I^{\rm obs}$, averaged over the area A:
\begin{equation} 
I^{\rm obs} = {1\over A}~ \int_{0}^{fA} I_0 e^{-X\tau} da +{1\over A}~ 
\int_{fA}^{A} I_0 e^{-\tau} da.
\end{equation}
Carrying out the integration and denoting
\begin{equation}  
\alpha_{\rm eff} \equiv \frac{\tau_{{\rm eff},V}}{\tau_V} \alpha  
\end{equation}
where,
\begin{equation}
 \tau_{{\rm eff},V} = -\ln(I^{\rm obs} / I_0) =
-\ln(fe^{-X\tau} + (1-f)e^{-\tau})
\end{equation}
 gives $\alpha_{\rm eff}$.  In Eq. 6 $\tau$ is the mean V extinction
, equal to 0.34, derived in \S 3.1.  If $f$ is set to zero
in Eq. 6 then the measured, or effective, extinction will be equal to
the mean.  Qualitatively the situation differs most 
for $f\sim 0.5$, where we obtain 
\begin{equation}
 \alpha_{\rm eff} = \frac {0.496}{0.34} 2.15 = 3.14,
\end{equation}
for a ratio of optical depths, X $\sim 2$.
Very similar estimates for $\tau_V / (\tau_I-\tau_H)$ have been
derived empirically by B97 (their Figure 3).
The histogram for $\tau_V = \alpha_{eff}(\tau_I-\tau_H)$, the clumpy extinction case, 
is shown by the dotted lines in Figure 2.  
The mean extinction in V has increased from 0.34 to 0.41, with a commensurate increase
of the dispersion.  Note, that the mean has changed by
more than the measurement errors.  Therefore, the assumed extinction curve
is the dominant of uncertainty in our analysis.

\section{Discussion}
\subsection{Comparison With Other Authors}

Optical depths through galaxy disks have been derived several ways by
different authors (Keel 1983, Keel and White 1995, James and Puxley 1995,
B97). Each analysis had to rely on different
sets of assumption, such as intrinsic galaxy symmetry (WK92, B97),
{\it case~B} recombination ratios for Balmer lines (James and Puxley 1995),
and intrinsic near-IR colors of galaxies (this paper). It is interesting to 
explore to which extent the results are impacted by the different techniques.

James and Puxley (1995) found quite high extinction values from analyzing
H$\beta$/H$\alpha$ ratios, seen though the center of NGC3314.
These values could be due to the small impact parameters of the probed
lines of sight, or because the extinction intrinsic to the 
HII regions has been under-corrected.

WK92 and B97 have estimated extinction by broad-band
photometric methods, WK92 using CCD data, B97 using data that
extend to $2.2\mu$m.  Referring all extinction values to face-on and the 
B-band, we find a mean of $\langle \tau_B\rangle \sim 0.55$ for UGC~5041
at $R\sim 1.4R_{exp}\sim 0.4R_{25}$ using our clumpy model, which
has an effective extinction curve similar to B97.  Keel and White (1995) 
list $\tau_B=0.1-0.4$ at somewhat
larger radii, and B97 find $\tau_B=0.6$ in an inter-arm region 
at $R\approx 0.7R_{25}$. B97 also find $\tau_B=1.3$ in the dust-lanes
near the arms, for which detailed case studies of nearby galaxies
(\eg Elmegreen 1980 and Rix and Rieke 1993) have already shown that
$\tau_B\gtorder 1$.

Even though we have analyzed a purely infrared-selected pair of galaxies,
and have based our analysis on near-IR data, it appears that our results
are quite consistent with the optical analyses, implying that 
the basic result of semi-transparent disks is not simply a
selection effect.

\subsection{Implications For Cosmologically Distant Objects}

As mentioned above, dust in foreground galaxies will dim and redden
distant background objects, such as QSOs, that appear in close projection.
On the other hand, the mass concentration
associated with the intervening galaxies may brighten the
background source through gravitational lensing 
(Hutchings 1990; Bartelmann et. al. 1996).  Hence the observed
incidence rate of close alignments between background sources 
and foreground galaxies depends on the competition between
the two effects.

Here we show how to use $p(\tau_V)$
to estimate the first effect, focusing on optically selected
quasars that lie within $\sim$ $( 5 to 10)$~kpc of galaxies, such as
QSOs lensed by spiral galaxies, or QSOs with low-redshift
damped Ly-$\alpha$ absorbers.  The radial range( $6.4$ to $10.3)$~kpc covered
by our estimate of p$(\tau_V)$ is quite limited, but of particular interest.  
First, because damped Ly-$\alpha$ absorbers have sizes $\sim10$~kpc (Briggs et. al. 1989).
We take from Peterson (1997, p.~174) the integrated
number--magnitude relation for quasars brighter than $M_l$: 
\begin{equation}
N(<M_l) = N_0 ~ 10^{0.8M_l} = N_0 ~ e^{1.84M_l}
\end{equation}
For objects whose light passes through a foreground galaxy, the
magnitude limit of any survey is effectively $M_l^{'}= M_l -
1.086\tau$, where $\tau$ is the optical depth at the wavelength of the
survey.  Given a probability distribution $p(\tau)$ we would expect such
sources to be under-represented by a factor of:
\begin{equation}
 {{N_{\rm obs}(<M_l)}\over{N(<M_l)}} = {1\over{N_0 e^{1.84M_l}}}
 \int_0^\infty N_0 e^{(M_l-1.08\tau)}~p(\tau)~d\tau.
\end{equation}
As Figure~3 shows $p(\tau)$ is well approximated by
\begin{equation} 
p(\tau) = \frac{1}{\sqrt{2\pi}\sigma_\tau}e^{-\frac{ (\tau - \langle\tau\rangle)^2}
{2\sigma^2_\tau}},
\end{equation}
which leads to
\begin{equation}
\frac {N_{\rm obs}(<M_l)}{N(<M_l)} =
e^{-1.98\langle\tau\rangle\bigl(1-\frac{\sigma_\tau^2}{\langle\tau\rangle}\bigr )}.
\end{equation}
For the derived values of $\langle\tau\rangle=0.34$ and $\sigma_\tau=0.15$,
and $\langle\tau\rangle=0.41$, $\sigma_\tau=0.19$, this results in
$\frac {N_{\rm obs}(<M_l)}{N(<M_l)} = 0.53$ 
and $\frac {N_{\rm obs}(<M_l)}{N(<M_l)}
= 0.47$, respectively.

Hence the chance of finding lensed of damped Ly-$\alpha$
absorbed quasars (with $m_B\ltorder 19$)
in such close projection ($\ltorder 10$~kpc) to spiral disks
is suppressed by about a factor of two due to dust extinction,
in good agreement 
with the independent estimate based on the Milky Way gas-to-dust ratio
by Perna, Loeb and Bartelmann (1997).

\section{Conclusion}

We have presented and analyzed H, I and B band images of the projected
galaxy pair UGC~5041, in order to explore the distribution of dust
extinction that light suffers while crossing 
the highly inclined foreground galaxy disk.
This galaxy pair is completely obscured in the optical,
and hence could have lead to $\langle \tau_V\rangle\gtorder 1$;
this system also provides 
optical depth constraints at quite small impact parameters.

After removing the light contribution from the foreground galaxy,
we compared, pixel by pixel, the observed I-H color of the (extincted)
background galaxy to its expected intrinsic I-H color, which can be
estimated to $\pm 0.08$~mag. This comparison resulted in a direct estimate
of the distribution of $~\tau_I - \tau_H$, with
$\langle \tau_I - \tau_H \rangle/\cos{72^\circ} = 0.6$.
For a homogeneous dust distribution this translates into
a face-on V-band optical depth of $\langle \tau_V\rangle =0.34$,
for clumpy dust this corresponds to $\langle \tau_V \rangle = 0.41$,
with dispersion of $0.15$ and $0.20$ in $\tau_V$.

Despite the different selection criteria (which would have permitted to
arrive at much higher optical depths) and an analysis based solely on
near-IR data, these optical depths are consistent with the
values found by WK92 and B97, and corroborate that 
generically $\tau^{face-on}_V\ltorder 1$ in nearby galaxy disks.

Using the derived distribution of $p(\tau_V)$, we estimated by 
how much quasars in close projection  ($5-10$~kpc) are under-represented
in optically selected, flux-limited samples ($m_{limit}(B) < 19$), 
and found a deficit by a factor of two due to dust extinction.

As in many other analyses, the limited spatial resolution,
which leads to averaging over areas of differing extinction
is an important limitation. We are
currently analyzing HST images of UGC5041 to repeat an analogous
analysis at vastly improved spatial resolution.

We would like to thank Gary Bernstein for providing the CCD frames 
UGC5041.  HWR was supported by an Alfred P. Sloan Fellowship.

\newpage
\clearpage

\begin{figure}
\epsscale{1.00}
\epsfxsize=\hsize\epsffile{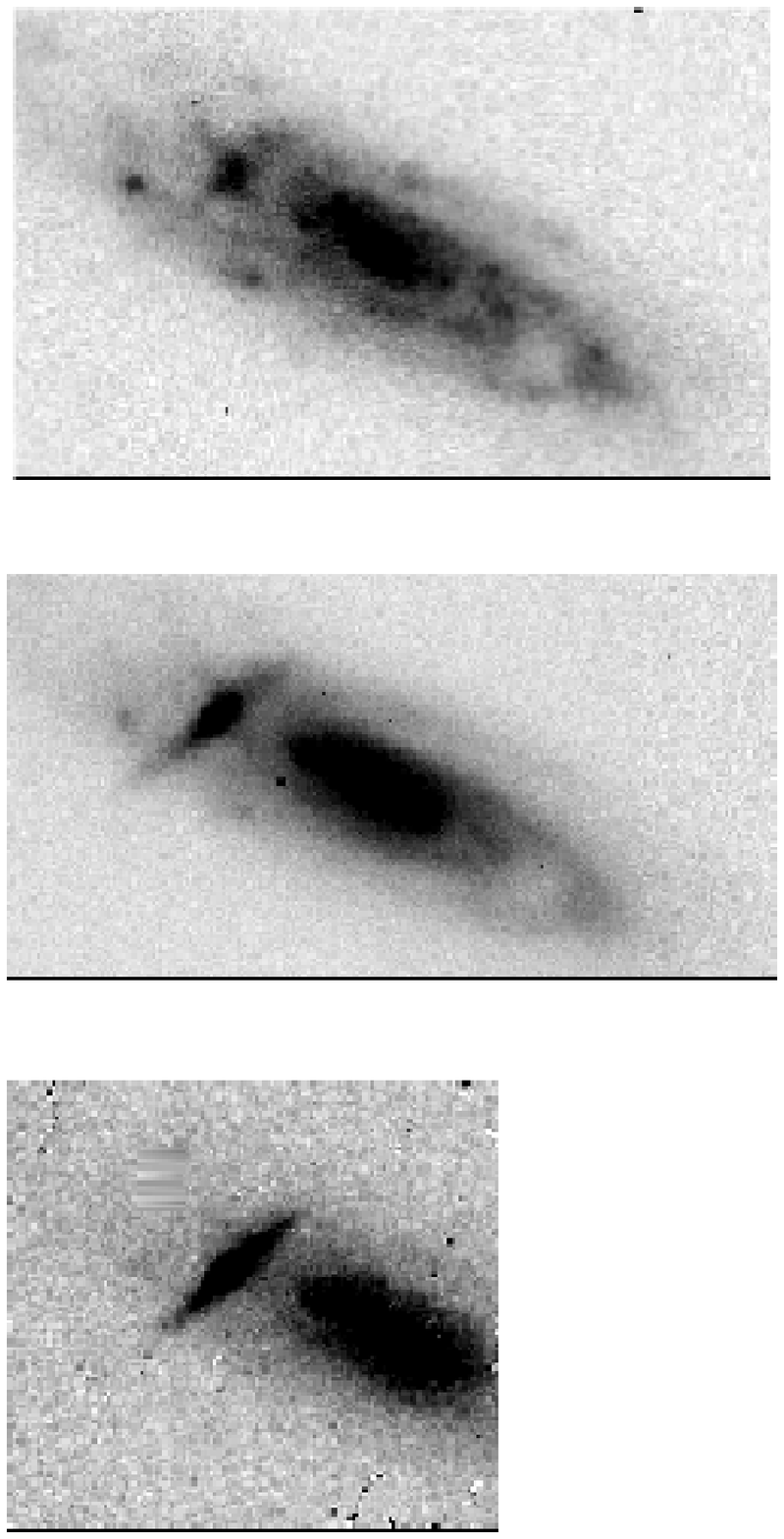}
\caption{ The B, I, and H band
images of the galaxy pair UGC5041, from top to bottom. 
 The center of the two galaxies are separated by $14.4''$.  
The images cover $39'' \times 64''$, and $35'' \times 66''$, 
and $38'' \times 53''$ in B, I, and H, respectively.}
\end{figure}
\clearpage

\begin{figure}
\epsscale{1.00}
\epsfxsize=\hsize\epsffile{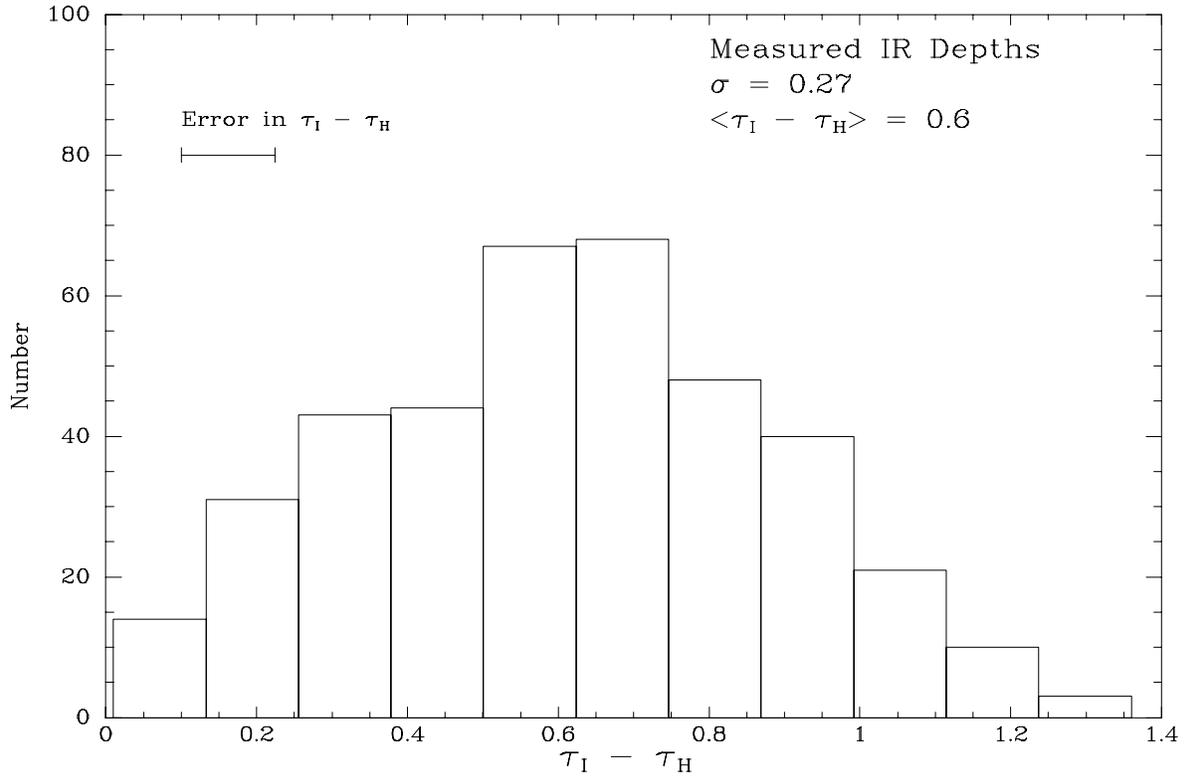}
\caption{ The pixel-by-pixel histogram of 
the differential optical depth $\tau_I-\tau_H$.  
The error ($\pm0.124$) is discussed in Section 2.2.  The distribution
can be approximated by a Gaussian with $<\tau_{I} - \tau_{H}> = 0.6$
and $\sigma_{\tau} = 0.27$.}
\end{figure}
\clearpage

\begin{figure}
\epsscale{1.00}
\epsfxsize=\hsize\epsffile{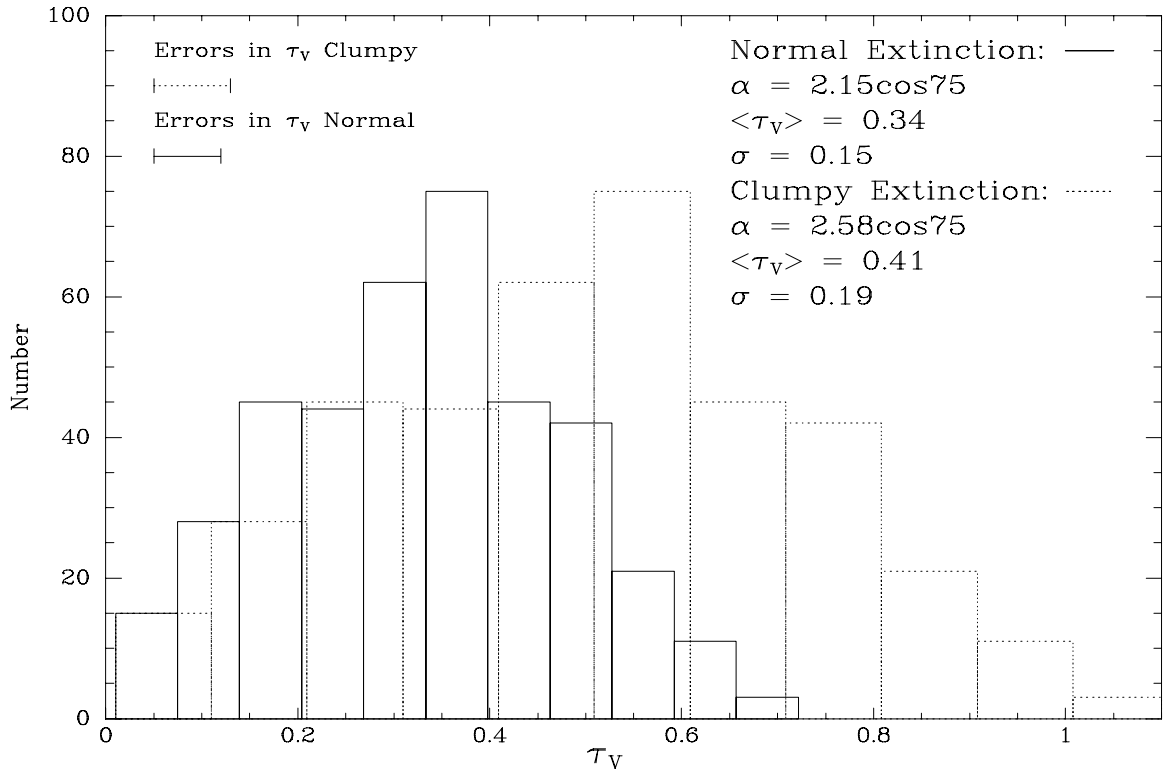}
\caption { Histogram of the face-on optical depth in V, 
assuming a homogeneous dust distribution (solid line) 
and a clumpy dust model (dashed line).  The errors 
are 0.07 and 0.08, respectively.}
\end{figure}
\clearpage

\end{document}